\documentclass[sigconf]{acmart}

\usepackage{ dsfont }
\usepackage{multirow}
\usepackage{enumitem}

\AtBeginDocument{%
  \providecommand\BibTeX{{%
    \normalfont B\kern-0.5em{\scshape i\kern-0.25em b}\kern-0.8em\TeX}}}

\setcopyright{acmcopyright}
\copyrightyear{2020}
\acmYear{2020}
\acmDOI{10.1145/1122445.1122456}

\acmConference[Xi'an '20]{Xi'an '20: ACM SIGIR Conference on Research and Development in Information Retrieval}{June 03--05, 2018}{Xi'an, China}
\acmBooktitle{Xi'an '20: ACM Symposium on Neural Gaze Detection,
  July 25--30, 20, Xi'an, China}
\acmPrice{15.00}
\acmISBN{978-1-4503-XXXX-X/18/06}

\begin{document}


\title{Learning Term Discrimination}

\author{Jibril Frej}
\email{jibril.frej@univ-grenoble-alpes.fr}
\affiliation{%
  \institution{Univ. Grenoble Alpes, CNRS, Grenoble INP*, LIG\\
* Institute of Engineering Univ. Grenoble Alpes}
}
\author{Philippe Mulhem}
\email{philippe.mulhem@imag.fr}
\affiliation{%
  \institution{Univ. Grenoble Alpes, CNRS, Grenoble INP*, LIG\\
* Institute of Engineering Univ. Grenoble Alpes}
}

\author{Didier Schwab}
\email{schwabd@univ-grenoble-alpes.fr}
\affiliation{%
  \institution{Univ. Grenoble Alpes, CNRS, Grenoble INP*, LIG\\
* Institute of Engineering Univ. Grenoble Alpes}
}

\author{Jean-Pierre Chevallet}
\email{jean-pierre.chevallet@univ-grenoble-alpes.fr}
\affiliation{%
  \institution{Univ. Grenoble Alpes, CNRS, Grenoble INP*, LIG\\
* Institute of Engineering Univ. Grenoble Alpes}
}

\renewcommand{\shortauthors}{Frej, et al.}

\begin{abstract}
    Document indexing is a key component for efficient information retrieval (IR). After preprocessing steps such as stemming and stop-word removal, document indexes usually store term-frequencies (tf). Along with tf (that only reflects the importance of a term in a document), traditional IR models use term discrimination values (TDVs) such as inverse document frequency (idf) to favor discriminative terms during retrieval. In this work, we propose to learn TDVs for document indexing with shallow neural networks that approximate traditional IR ranking functions such as TF-IDF and BM25. Our proposal outperforms, both in terms of nDCG and recall, traditional approaches, even with few positively labelled query-document pairs as learning data. Our learned TDVs, when used to filter out terms of the vocabulary that have zero discrimination value, allow to both significantly lower the memory footprint of the inverted index and speed up the retrieval process (BM25 is up to 3~times faster), without degrading retrieval quality.
\end{abstract}


\maketitle

\section{Introduction}

Document indexing for information retrieval (IR) usually consists in associating each document of a collection with a set of weighted terms reflecting its information content. To this end, a term discrimination value (TDV) is used to represent the usefulness of a term as a discriminator among documents~\cite{DBLP:books/daglib/0067811}. However, traditional IR systems make little use of TDVs during indexing. The only exception is stopword removal which considers that stop-words have null discrimination value and removes them from document representations. Stop-word removal also speeds up the retrieval process when using an inverted index since it removes stop-words that have long posting lists.

\noindent \textbf{Related Work.} Several methods have been proposed to compute TDVs, such as using the density of the document vector space~\cite{DBLP:journals/cacm/SaltonWY75} or the covering coefficient of documents~\cite{DBLP:journals/jasis/CanO87}. Nowadays, the most common approaches in traditional IR models for computing TDVs are to use either the inverted document frequency (idf)~\cite{DBLP:journals/ftir/RobertsonZ09} or a smoothing method such as Bayesian smoothing using Dirichlet prior~\cite{DBLP:conf/sigir/ZhaiL01}. Recently, Roy et al.~\cite{DBLP:conf/sigir/RoyBM19} proposed to select discriminative terms to enhance query expansion methods based on pseudo-relevance feedback. However, these approaches use TDVs only at retrieval time and not during indexation. Inspired by stop-word removal, we suggest that using supervised learning to remove non discriminative terms at indexation can speed up the retrieval process with no deterioration of retrieval quality.

\noindent \textbf{Our Contributions.} In this work, we propose to learn TDVs in a supervised setting using a shallow neural network and word embeddings. In order to have TDVs adapted to traditional IR ranking functions, we propose to learn TDVs by optimizing the ranking of traditional IR models. However, components of these models such as term frequency (tf) or inverse document frequency (idf) are not differentiable in the setting in which neural networks are commonly used (sequences of word embeddings processed by CNN, RNN or Transformer Layers). This non-differentiability makes impossible the use of gradient descent-based optimization methods required by neural networks. Hence, we propose a setting that uses bag of words (BoWs) as sparse vectors to have differentiable tf and $\ell_1$-norm as an approximation to the $\ell_0$-norm to have a differentiable approximation of the idf. Hence, we learn TDVs to optimize differentiable approximations of traditional IR ranking functions. Since we are using a shallow neural network with few parameters, our models do not need large amounts of positively labelled query-documents pairs to outperform traditional IR models. Additionally, we remove posting lists associated with terms having zero TDV from the inverted index in order to significantly enhance retrieval speed.

\noindent In short, our contributions are : 
\begin{itemize}[noitemsep,nolistsep]
    \item A new framework for differentiable traditional IR;
    \item Differentiable versions of IR functions to learn TDVs;
    \item A significant speed up retrieval obtained by removing posting lists associated to terms with zero TDV;
\end{itemize}

\section{Learning term discrimination}

To learn TDVs adapted to traditional IR ranking functions using neural networks, we propose the following strategy: 
\begin{enumerate}[noitemsep,nolistsep]
    \item Make traditional IR ranking functions compatible with neural networks by using matrix operations that are differentiable with respect to the inverted index (Section~\ref{subsec:matrix_view});
    \item Introduce a shallow neural network to compute TDVs and a method to include TDVs into the inverted index (Section~\ref{subsec:shallow});
    \item Use the differentiable functions proposed in Section~\ref{subsec:matrix_view} to learn TDVs adapted to traditional IR ranking functions using a supervised shallow neural network (Section~\ref{subsec:e2eTDV});
\end{enumerate}

\subsection{Differentiable traditional IR}
\label{subsec:matrix_view}
All operations used by traditional IR ranking function can be derived from the inverted index. Given a vocabulary $V$ and a collection $C$, the inverted index can be represented as a sparse matrix $S \in \mathds{R}^{|V|\times|C|}$. Each element of $S$ corresponds to the term frequency (tf) of a term $t \in V$ with respect to (w.r.t) a document $d \in C$: $S_{t,d} = \text{tf}_{td}$. Columns of $S$ (denoted as $S_{:,d}$) correspond to the BoW representations of documents in $C$ and rows of $S$ (denoted as $S_{t,:}$) correspond to the posting lists of terms in $V$. Let $Q\in\mathds{N}^{|V|}$ denote the BoW representation of a query $q$.

\subsubsection{TF-IDF} 
Using matrix operations over $S$, the TF-IDF ranking function between a query $q$ and a document $d$ can be formulated as:

\begin{align}
\text{TF-IDF}(q,d) = \sum\limits_{t \in q} \text{tf}_{td}\; \text{idf}_t = Q^\intercal\cdot \left(S_{:,d} \odot \text{IDF} \right),
\label{eq:tdv_tf}
\end{align}
where $\odot$ denotes the element wise (or Hadamard) product and $\text{IDF}\in \mathds{R}^{|V|}$ denotes the vector containing inverse document frequencies (idf) of all terms $V$. idf can be derived from $S$ using the $\ell_0$-norm to compute document frequencies (df):
\begin{equation}
\text{idf}_t = \log\frac{|C|+1}{\text{df}_t} = \log\frac{|C|+1}{\ell_0(S_{t,:})}.
\end{equation}

To be able to have TDVs adapted to traditional IR ranking functions, we want such functions to be differentiable w.r.t elements of $S$. However, the $\ell_0$-norm is non differentiable. Consequently, we propose to redefine idf using $\ell_1$ which is a good approximation to $\ell_0$~\cite{ramirez2013l1}. If we replace $\ell_0$ with $\ell_1$, the obtained idf will be negative for terms such that $\ell_0(S_{t,:}) > |C| + 1$ which would violate the Term Frequency Constraint, a desirable property of retrieval formula~\cite{DBLP:conf/sigir/FangTZ04}. To ensure positives idfs, we propose a maximum \ normalization:
\begin{equation}
\widetilde{\text{idf}_t} = \log\frac{\max_{\{ t'\in V\}}\ell_1(S_{t',:})+1}{\ell_1(S_{t,:})}.
\end{equation}

Using $\widetilde{\text{idf}_t}$, we have the following differentiable approximation of the TF-IDF formula, denoted as $\widetilde{\text{TF-IDF}}$:

\begin{align}
\widetilde{\text{TF-IDF}}(q,d) &= Q^\intercal\cdot \left(S_{:,d}\odot \widetilde{\text{IDF}}\right) = \sum\limits_{t \in q} S_{t,d} \; \widetilde{\text{idf}_t}.
\label{eq:tf_idf}
\end{align}
where $\widetilde{\text{IDF}}\in \mathds{R}^{|V|}$ is the vector containing $\widetilde{\text{idf}_t}$ of all terms in $V$.

\subsubsection{BM25}


We can also define a differentiable approximation of the BM25 ranking formula using $\widetilde{\text{IDF}}$:

\begin{align}
\widetilde{\text{BM25}}(q,d) = Q^\intercal\cdot \widetilde{\text{IDF}} &\odot \left(S_{:,d} \left(k_1 + 1 \right)  \right)  \nonumber \\
&./ \left( S_{:,d} + k_1\left( 1 - b + b\frac{|d|}{avgdl} \right) \mathds{1}_{|V|} \right),
\label{eq:BM25}
\end{align}

where $k_1$ and $b$ are parameters of BM25 and $avgdl$ denotes the average length of documents in $C$. $./$  is an element wise (or Hadamard) division and $\mathds{1}_{|V|}$ is a vector of dimension $|V|$ with all elements equal to one. Both $|d|$ and $avgdl$ are differentiable w.r.t the elements of $S$: $|d| = \ell_1(S_{:,d})$ and $avgdl = \sum\limits_{d\in C}\ell_1(S_{:,d})/|C|$.

\subsubsection{Dirichlet Language Model}
We also propose a differentiable language model with Dirichlet prior smoothing~\cite{DBLP:conf/sigir/ZhaiL01}:

\begin{align}
    \text{LM}(q,d) &= \sum\limits_{t\in q} \log \left( 1 + \frac{\text{tf}_{td}}{\mu p(t|C)}\right) + |q| \log\alpha_d \nonumber\\
    &= Q^\intercal\cdot [ \log\left(\mathds{1}_{|V|} +  \left( S_{:,d} ./ (\mu P_C) \right) \right)\nonumber\\
    &+ |q|\log(\alpha_d) \mathds{1}_{|V|}],
    \label{eq:lm}
\end{align}

where $\mu$ is a parameter of LM, $\alpha_d = \frac{\mu}{|d| + \mu}$ is a document dependent constant and $P_C\in \mathds{R}^{|V|}$ is the vector containing the probability of a term given the collection language models for all terms in the vocabulary: $ \forall t \in V, P_{C_t} = p(t|C) = \sum\limits_{d \in C} S_{t,d} / \sum\limits_{t' \in V}\sum\limits_{d \in C} S_{t',d}$.

\subsection{Shallow neural network for learning TDVs}
\label{subsec:shallow}
To have a model that requires few training data, we propose to compute the TDVs using a shallow neural network composed of a single linear layer and the Rectified Linear Unit (ReLU) non linearity: $\text{tdv}_t = \text{ReLU}(w_t^\intercal \cdot w +b) = \max(0,w_t^\intercal\cdot w +b)$ where $w_t$ is the word embedding of term $t$, and $w$ and $b$ are parameters of the neural network. We employ ReLU activation function to ensure that the TDV is positive (as negative TDVs can violate the Term Frequency Constrain) and to be able to have terms with zero TDV that we can remove from the inverted index. We redefine the inverted index $S$ using TDVs the following way: 

\begin{equation}
    S'_{t,d} = \text{tf}_{td} \; \text{tdv}_{t} = \text{tf}_{td} \; \text{ReLU}(w_t^\intercal \cdot w +b).
\end{equation}
With this definition, we ensure that if a term $t$ has zero discrimination value ($\text{tdv}_{t} = 0$), the row in $S$ associated to $t$ is filled with zeros, therefore it's posting list is empty and can be removed from the inverted index.

\subsection{Learning TDVs with differentiable IR}
\label{subsec:e2eTDV}

\begin{figure}[ht]
  \centering
  \includegraphics[width=\linewidth]{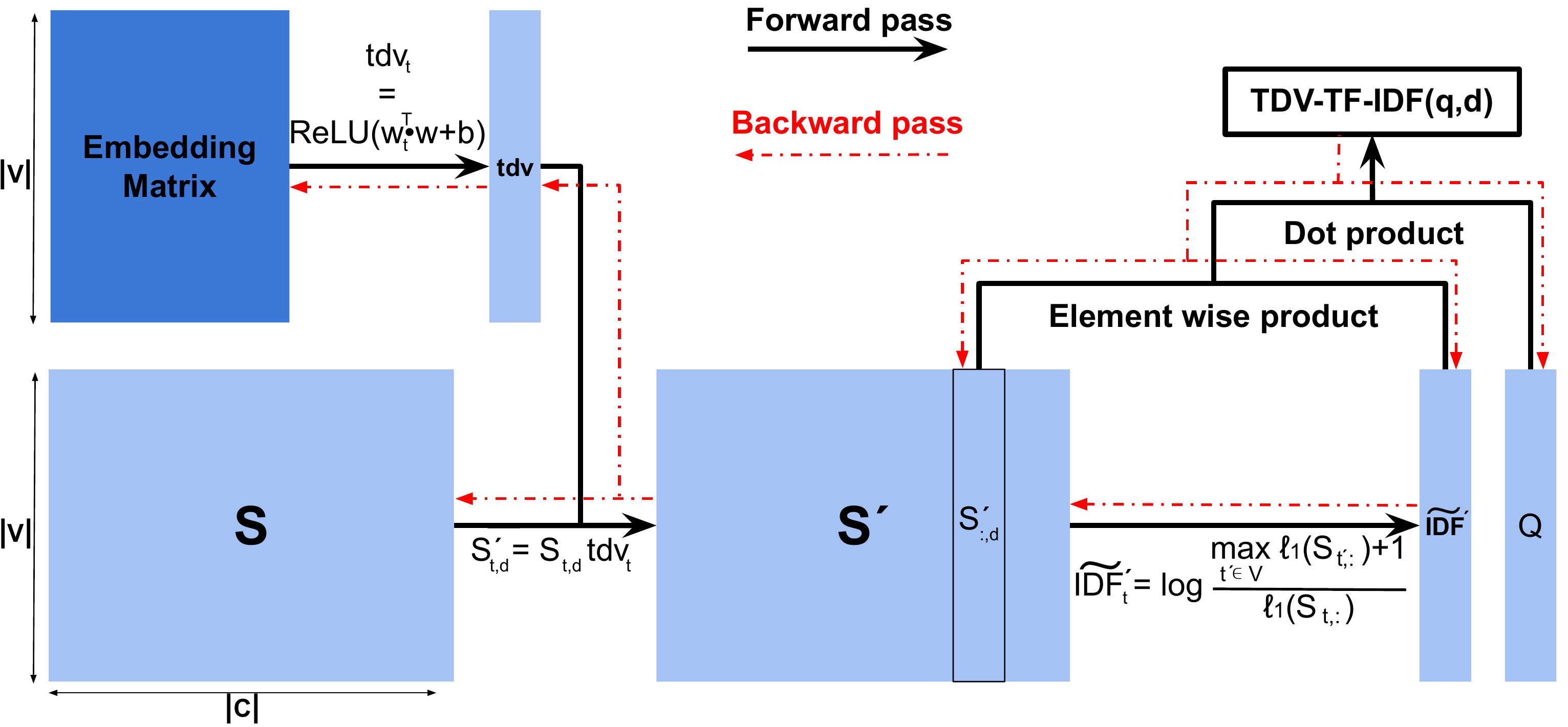}
  \caption{Architecture of TDV-TF-IDF. All operations are differentiable and gradients can be back-propagated from the final score to $w$ and $b$.}
  \label{fig:archi}
\end{figure}

To learn TDVs that optimize the score of traditional IR ranking formulae, we simply replace $S$ in Equations~(\ref{eq:tf_idf}),~(\ref{eq:BM25}) and (\ref{eq:lm}) by $S'$:

\begin{alignat}{2}
    &\text{TDV-TF-IDF}(q,d) &&= Q^\intercal\cdot \left(S'_{:,d}\odot \widetilde{\text{IDF}'}\right);\\
    &\text{TDV-BM25}(q,d) &&= Q^\intercal\cdot \widetilde{\text{IDF}'} \odot \left( S'_{:,d} \left(k_1 + 1 \right)\right)\nonumber\\
    & &&./ \left( S_{:,d} + k_1\left( 1 - b + b\frac{|d|'}{avgdl'} \right) \mathds{1}_{|V|} \right);\\
    &\text{TDV-LM}(q,d) &&= Q^\intercal\cdot [ \log ( \mathds{1}_{|V|} + ( S'_{:,d} ./ (\mu P'_C) ) )\nonumber\\
    & &&+  |q|\log(\alpha'_d) \mathds{1}_{|V|}];
\end{alignat}

\noindent where $\widetilde{\text{IDF}'}$, $|d|'$, $avgdl'$ and $\alpha'_d$ denote respectively $\widetilde{\text{IDF}}$, $|d|$, $avgdl$ and $\alpha_d$ computed with $S'$ instead of $S$. Scores computed by these ranking functions are differentiable w.r.t parameters $w$ and $b$ (see Figure~\ref{fig:archi}). Consequently, we can use gradient descent-based optimization methods to update $w$ and $b$ in order to compute TDVs adapted to traditional IR ranking functions.

\subsection{Training}

We use the pairwise hinge loss function as our ranking objective:
\begin{equation}
    \mathcal{L}_{\text{Hinge}}(f,q,d^+,d^-) = \max(0,1-f(q,d^+) + f(q,d^-)),
\end{equation}
where $f$ is a differentiable ranking function for IR and $d^+$ is a document more relevant than $d^-$ w.r.t query $q$. To ensure that the ranking functions produce terms with zero discrimination values, we also use a sparsity objective during training. To do so, we minimize the $\ell_1$-norm of the document BOWs representation as suggested by Zamani et al.~\cite{DBLP:conf/cikm/ZamaniDCLK18}. The final loss function is defined as follows: 

\begin{equation}
    (1-\lambda)\mathcal{L}_{\text{Hinge}}(f,q,d^+,d^-) + \lambda( \ell_1(Sf'_{:,d^+}) + \ell_1(Sf'_{:,d^-})),
\end{equation}
where $\lambda \in [0,1]$ is the regularization hyper-parameter and $Sf'$ is the inverted index matrix computed by $f$.

\section{Experiments}

\subsection{Collections}
As mentioned previously, our models need few positively labelled query-documents pairs (qrels). Consequently, we evaluated them on 3 standard TREC collections : 
\begin{itemize}[noitemsep,nolistsep]
  \item AP88-89 with topics 51-200 and 15 856 positive qrels
  \item FT91-94 with topics 251-450 and 6 486 positive qrels
  \item LA with topics 301-450 and 3 535 positive qrels
\end{itemize}

\noindent We use title of topics as queries. We lowercase, stem and remove stop-words from collections.

\subsection{Baselines}
We compare our models with several traditional IR models using a standard inverted index: TF-IDF, LM and BM25 and with neural supervised approaches for IR: DRMM~\cite{DBLP:conf/cikm/GuoFAC16}, DUET~\cite{DBLP:conf/www/Mitra0C17} and Conv-KNRM~\cite{DBLP:conf/wsdm/DaiXC018}. 
DRMM performs matching based on a histogram of cosine similarities between word embeddings of query and document. The DUET model is a deep architecture that uses both local (exact matching signal) and distributed (word embeddings) representations to compute a relevance score. Conv-KNRM uses convolutions to generate several query-document interaction matrices that are processed by kernel pooling to produce learning-to-rank features.

\subsection{Implementation}
\textbf{\label{sec:implementation}}
We implemented and trained our models with \textit{Tensorflow}~\cite{DBLP:journals/corr/AbadiABBCCCDDDG16}. We used \textit{MatchZoo}~\cite{DBLP:conf/sigir/GuoFJC19} for training and evaluation of neural baselines. We implemented traditional IR ranking functions in \textit{Python}. We used word embeddings pre-trained on Wikipedia with the \textit{fastText}~\cite{DBLP:conf/lrec/MikolovGBPJ18} algorithm. Because of the limited amount of training data, we did not fine-tune word embeddings. In order to ensure that TDVs for all terms are non zero at the beginning of the training, we initialized  bias $b$ to 1. Weight vector $w$ is  initialized with the default \textit{Tensorflow} initialization. To accelerate the training process, collection-level measures such as idf and collection language models were implemented batch wise and not collection wise. Preliminary experiments showed that dropout does not allow for better performance on the validations sets which is probably due to the low number of parameters of our models. Therefore, we do not use dropout in our experiments. 5-fold cross validation across the queries of the collections is used to tune hyperparameters. We use the Adam~\cite{DBLP:journals/corr/KingmaB14} algorithm to optimize models and early stopping on the training nDCG@5.

\begin{table}[ht]
    \centering
    \begin{tabular}{@{\extracolsep{1pt}}lcccccc@{}}
    \hline
    \multirow{ 2}{*}{Method} & \multicolumn{2}{c}{AP88-89} & \multicolumn{2}{c}{LA} & \multicolumn{2}{c}{FT91-94} \\
    \cline{2-3} \cline{4-5} \cline{6-7}
        & Base. & TDV & Base. & TDV & Base. & TDV\\
    \hline
    TF-IDF  & 147.4 & 29.2  & 26.3  & 4.6  & 56.6  & 15.1\\
    LM      & 683.8 & 180.1 & 139.6 & 54.3 & 270.9 & 122.6\\
    BM25    & 207.0 & 61.2  & 29.2  & 16.6 & 83.1  & 42.8\\
    DRMM    & >$10^4$ & \textbackslash  & >$10^4$  & \textbackslash & >$10^4$  & \textbackslash\\
    DUET    & >$10^5$ & \textbackslash  & >$10^5$  & \textbackslash & >$10^5$  & \textbackslash\\
    Conv-KNRM    & >$10^5$ & \textbackslash  & >$10^5$  & \textbackslash & >$10^5$  & \textbackslash\\
    \hline
    \end{tabular}
    \caption{Comparison of the average retrieval time per query in milliseconds.}
    \label{table:retr_time}
\end{table}

\begin{table*}
    \centering
    \begin{tabular}{@{\extracolsep{1pt}}lll|ll|ll|ll|ll|ll@{}}
        \hline
        \multirow{ 2}{*}{Method} & \multicolumn{4}{c}{AP88-89} & \multicolumn{4}{c}{LA} & \multicolumn{4}{c}{FT91-94} \\
        \cline{2-5} \cline{6-9} \cline{10-13}
        & \multicolumn{2}{c}{nDCG@5} & \multicolumn{2}{c}{Recall@1000} &\multicolumn{2}{c}{nDCG@5} & \multicolumn{2}{c}{Recall@1000} & \multicolumn{2}{c}{nDCG@5} & \multicolumn{2}{c}{Recall@1000}\\
        & Baseline & TDV & Baseline & TDV & Baseline & TDV & Baseline & TDV & Baseline & TDV & Baseline & TDV \\ 
        \hline
        TF-IDF      & 26.38 & 30.28$^*$ & 53.08 & 58.42$^*$ & 17.92 & 23.54$^*$ & 60.05 & 65.09$^*$ & 17.82 & 25.11$^*$ & 50.98 & 55.55$^*$\\
        LM          & 44.64 & 46.30$^*$ & 67.26 & 67.98 & 34.50 & 36.16 & 69.15 & 70.29 & 35.15 & \textbf{37.78$^*$} & 59.63 & 61.56\\
        BM25        & 44.70 & \textbf{47.09$^*$} & 67.09 & 66.78 & 34.98 & \textbf{40.04$^*$} & 68.47 & \textbf{72.00$^*$} & 35.31 & 36.98$^*$ & 60.40 & \textbf{62.68}\\
        DRMM        & 44.05 & \textbackslash & \textbf{68.24} & \textbackslash & 36.22 & \textbackslash & 70.41 & \textbackslash & 35.95 & \textbackslash & 61.42 & \textbackslash\\
        DUET        & 43.86 & \textbackslash & 67.86 & \textbackslash & 15.26 & \textbackslash & 58.65 & \textbackslash & 16.88 & \textbackslash & 45.25 & \textbackslash\\
        Conv-KNRM        & 44.13 & \textbackslash & 67.36 & \textbackslash & 22.65 & \textbackslash & 60.41 & \textbackslash & 37.95 & \textbackslash & 51.42 & \textbackslash\\
        \hline
    \end{tabular}
    \caption{Performance comparison of the proposed models and baselines. Best results for each metric on each collection are highlighted in bold. * indicates a statistically significant improvement (p<0.05) of TDV over baselines. DRMM, DUET and Conv-KNRM do not outperform statistically significantly TDV-BM25 or TDV-LM.}
    \label{table:perf}
\end{table*}

\begin{table}[h]
    \centering
    \begin{tabular}{lccc}
    
        \hline
        Method & AP88-89 & LA & FT91-94 \\
        \hline
        TDV-TF-IDF  & -45.00\% & -39.67\% & -45.77\%\\
        TDV-LM      & -46.06\% & -34.03\% & -40.25\%\\
        TDV-BM25    & -46.91\% & -32.35\% & -44.42\%\\
        \hline
    \end{tabular}
    \caption{Inverted index memory footprint reduction.}
    \label{table:removed}
\end{table}

\subsection{Evaluation}
We assess three different evaluation measures: (1) standard IR metrics: nDCG@5 and Recall@1000; (2) inverted index's memory footprint reduction after removing terms with zero discrimination value; (3) average retrieval time per query. Statistically significant differences of nDCG@5 and Recall@1000  are computed with the two-tailed paired t-test with Bonferroni correction.
\section{Results}

\noindent\textbf{Retrieval speed up.} Table~(\ref{table:retr_time}) reports the retrieval time of the different approaches. First, we notice that the neural baselines are dramatically slower at retrieving documents than other models that use inverted indexes. Second, by filtering out terms with zero discrimination value from the inverted index, we are able to significantly speed up the retrieval process of all ranking functions on all collections. On LA and FT91-94, BM25 retrieval speed is almost doubled and on AP88-89 BM25 retrieval speed is tripled. Interestingly, we notice that LM consistently takes more time to retrieve documents than other models. Indeed, the ranking formula of LM has to compute logarithms (that are computationally expensive) at retrieval time whereas BM25 and TF-IDF can compute such operations during indexation and use a lookup table at retrieval. 

\noindent\textbf{Effectiveness of proposed approaches.} Table~(\ref{table:perf}) shows the performance comparison of baselines and our models. Our main observation is that in most cases, learning and incorporating TDVs into IR ranking functions improves their performances despite the small amount of training data. Indeed, by construction and as a results of the bias initialization described in Section~(\ref{sec:implementation}), our models performance are already close to the baselines at the start of the training process. Moreover, since we are in a limited data scenario, neural baselines perform poorly on the TREC collections compared to the traditional ones. The exception being DRMM as it has few parameters and does not require large amount of training data.

\noindent\textbf{Memory footprint reduction.} Table~(\ref{table:removed}) describes the inverted index memory footprint reduction obtained when removing terms with zero discrimination value from the inverted index. For all collections, we are able to remove a significant portion of the inverted index and still outperform the original ranking functions. 

\section{Conclusion}
In this paper, we proposed to learn TDVs using supervised learning. In order to have TDVs specific to traditional IR ranking functions and to be able to use neural networks, we developed a framework to make such functions differentiable and compatible with matrix operations. Moreover, our models can be trained with few positively labelled data. Removing terms with zero TDV at indexation leads to drastic retrieval speed up with a slight performance improvement compared to BM25 on several TREC collections. As future work, we are studying the correlation between the learned TDVs and count based formulae such as idf. We also plan to evaluate our models on collections with larger amount of labelled data.

\bibliographystyle{ACM-Reference-Format}
\bibliography{sample-base}

\end{document}